# Bonding-aware Materials Representation for Deep Learning Atomistic Models

*Or Shafir and Ilya Grinberg*

*Department of Chemistry, Bar-Ilan University, Ramat-Gan, Israel*

**Abstract**

Deep potentials for molecular dynamics (MD) achieve first-principles accuracy at much lower computational cost. However, their use in large length- and time-scale simulations is limited by their lower speeds compared to analytical atomistic potentials, primarily due to network complexity and long embedding time. Here, based on the moments theorem, we develop a chemical-bonding-aware embedding for neural network potentials that achieve state-of-the-art accuracy in forces and local electronic density of states prediction with an ultrasmall 16×32 neural network resulting in significantly lower computational cost.

The evaluation of electronic structure and interatomic forces plays a crucial role in the study of materials. Traditionally, ab initio methods, such as density functional theory (DFT), have been used to investigate the electronic structure and interatomic forces of materials. However, DFT calculations are computationally expensive and time-consuming, which makes it challenging to study materials at large length scale or perform long MD simulations. While this issue can be addressed by using empirical atomistic potentials [1–3] in the simulations, the accuracy of such potentials is limited. Furthermore, due to the absence of electrons in atomistic models, atomistic potentials cannot model electronic structure changes during MD simulations.

Machine learning (ML) and deep learning (DL) have emerged as alternative approaches for studying materials [4–15]. ML and DL potentials can reproduce the accuracy of DFT for the potential energy surface of the data used in the training process. Various networks are employed for ML potentials, including the Behler-Parinello neural network (BPNN) [5,16], the SchNet framework [17,18], spectral neighbor analysis potentials [19,20], and others [21–26], with several libraries facilitating the development of these networks [27,28]. For the ML potential development, an embedding must be used (or learned while training). These embeddings capture the structural characteristics and chemical environment of atoms in a more sophisticated manner. While the naïve approach of using the systems' coordinates is valid for the development of deep potentials, it relies on the networks' ability to identify the patterns giving rise to the target property. Embeddings such as the atom-centered symmetry functions [29], Smooth Overlap of Atomic Positions (SOAP) [30], Many-Body Tensor Representation (MBTR) [31], Coulomb/Ewald summation matrix, and others, offer a more detailed and comprehensive descriptions of material properties compared to solely supplying the coordinates of the structure.

Despite their success, ML and DL algorithms also have several limitations such as reliance on large amounts of training and test data, which can be time-consuming and difficult to obtain. In addition, for the application of DL in MD simulations, the prediction times are the major challenge to overcome, with MD simulations using DL-based potentials typically slower by one or two orders of magnitude compared to the standard atomistic analytical potentials [14].

When working with large networks in materials science tasks, one of the major challenges is the incorporation of prior knowledge relevant to the specific problem at hand so as to avoid the need for the network to relearn the known physics and chemistry principles that influence the target property. While physics-informed neural networks have been developed to address this issue [32], they require a definition of an exact loss equation for the problem, which can be challenging for certain tasks. A potential solution for the incorporation of prior chemical and physical knowledge is to use embedding techniques, which offer a favorable approach due to their simplicity. By employing chemical-bonding-aware embedding, we can effectively integrate existing knowledge into the network's architecture and enhance its performance in capturing relevant features and patterns.

Here, we present a set of features based on the moments theorem of Cyrot-Lackmann and Ducastelle [33,34] that carries the information of the underlying chemistry and physics related to the target property. The $n$th moment ($\mu^n$) of the local density of states (LDOS) ($D(E)$) of the orbital $\alpha$ of atom $i$ is defined as

$$\mu_{i\alpha}^{(n)} = \int_{-\infty}^{+\infty} E^n D_{i\alpha}(E) dE \qquad \text{Equation 1}$$

where $E$ is the energy. The moments theorem states that in the tight-binding approximation, the $n$th moment of the local DOS can be calculated by summing the products of the Hamiltonian matrix elements along all self-returning paths of length $n$, beginning and ending at orbital $\alpha$ of atom $i$. For orbital $\alpha$ of atom $i$, the product along the $n$ steps path $i\alpha \rightarrow j_1\beta_1 \rightarrow j_2\beta_2 \rightarrow \cdots \rightarrow i\alpha$ which gives the $\mu_{i\alpha}^{(n)}$ DOS moment can be written

$$\mu_{i\alpha}^{(n)} = \sum_{j_1\beta_1\ldots j_{n-1}\beta_{n-1}} \langle i\alpha|\widehat{H}|j_1\beta_1\rangle \langle j_1\beta_1|\widehat{H}|j_2\beta_2\rangle \times \ldots \times \langle j_{n-1}\beta_{n-1}|\widehat{H}|i\alpha\rangle \qquad \text{Equation 2}$$

where all paths hops through nearest neighbors overlapping orbitals (namely $j\beta$), beginning and ending with the same atom and orbital (namely $i\alpha$). The length of the paths corresponds to the order of moment (namely $n$). This suggests that for a given atom, the electronic structure as characterized by LDOS and therefore the atom's forces and its contribution to the total energy are either controlled or at least strongly influenced by the products of the hopping matrix elements along the self-returning paths. Therefore, the embedding using cyclic paths provides information about the chemical bonding in the material, enabling accurate chemical description. Furthermore, the encapsulation of the information about the relationship between the atomic positions and chemical bonding in the embedding allows accurate modeling of electronic structure (LDOS) using atomistic information only.

We demonstrate high accuracy in the prediction of both forces and LDOS of $BaTiO_3$ and $MoO_3$ using the moments-based embedding and the ability to increase the prediction accuracy by systematically increasing the embedding complexity. Such chemical-bonding-relevant features allow a more efficient description of the system and, therefore, utilization of a simpler ML-based model, as demonstrated here with the application of an ultrasmall in modern terms (16×32) dense fully connected neural network which we call TinyNet. Our embedding uses a one-dimensional vector, which is calculated based on the moments theorem to represent each atom in the material (see SI Section I for details regarding the development of the representation). This vector captures the information about the self-repeating paths of bonds up to a certain length, denoted as $n$. It includes the negative exponent of the sum of interatomic distances between the bonded atoms along these paths, with each entry representing a different path. This part of the vector is denoted as $L_n$. By combining $L_n$ with the relative Cartesian coordinates of the nearest neighbors, we create a vector called $L_n^{(N)}$. For highly bonded structures, the $L_n^{(N)}$ vector is relatively large, and thereby can be simplified by only using paths without repetitions (i.e., simple cycles). The combination of addition of coordinates of neighbors and all self-returning paths to form $L_n^{(N)}$ provides a comprehensive description of the bonding and local interatomic interactions in the system.

This embedding was used to develop atomistic potentials for $MoO_3$ and $BaTiO_3$ datasets (800 and 1000 structures, respectively, see SI Section II for computational details) obtained using DFT calculations performed with the SIESTA code [35] and norm-conserving pseudopotentials from the PseudoDojo repository [36]. To develop the potentials for the prediction of the atomic forces and the LDOS, we train fully connected neural networks in the BPNN architecture [5], utilizing a varying number of layers and nodes as described below. The Tensorflow package [37] for Python was employed for this purpose. All networks were trained using a 80-20% train-test split with mean squared error (MSE) as the loss function.

We begin our analysis of the results with the prediction of forces for the MoO$_3$ system. To determine the appropriate number of hops to use for forces prediction, we evaluated the performance of our models using different values of $n$ and $N$. Figure 1 (a) displays the validation loss of the predicted total forces ($F_{tot}$) plotted versus $N$ for $L_n^{(N)}$, with varying combinations of $n$ after training for 100,000 epochs. Analysis of Figure 1 (a) reveals a consistent decrease in the error from 2 to 8 hops (i.e., $2 \leq n \leq 8$), ultimately reaching an MSE of 0.0086 (eV/Å)$^2$ for the validation set. This systematic reduction in loss with increasing $N$ can be attributed to the incorporation of additional information. Furthermore, in most cases, we observe a decrease in loss with higher values of $n$. However, when $n = 8$ is used, slight increases in the final loss are observed, which can be attributed to its higher dimensionality (see Figure S1 for vectors shape as a function of $n$ and $N$). We note that for the prediction of forces, the effect on the loss for larger number of hops $n$ is less significant than that of $N$, as suggested by the large MSE of the $L_0$ features (see Figure S2).

For further training, assessment and application, we select $L_6^{(4)}$ and $L_4^{(2)}$. The choice of $L_6^{(4)}$ is based on its performance, and the inclusion of $L_4^{(2)}$ in our analysis is due to its compact size and relatively high accuracy. Using the $L_6^{(4)}$ and $L_4^{(2)}$ features, we train different network architectures. We observed a consistent loss decrease with the increase in the number of nodes and layers for both features, as depicted in Figure 1 (b) when comparing different architectures, and in Figure S3 when considering the number of parameters. To achieve a fast prediction time and ensure feasibility across various applications, we opted for a simpler architecture. Therefore, we selected the second-smallest network, which had dimensions of 16×32 while still maintaining high accuracy. Figure S4 compares the embedding and prediction times for the different models based on the different features.

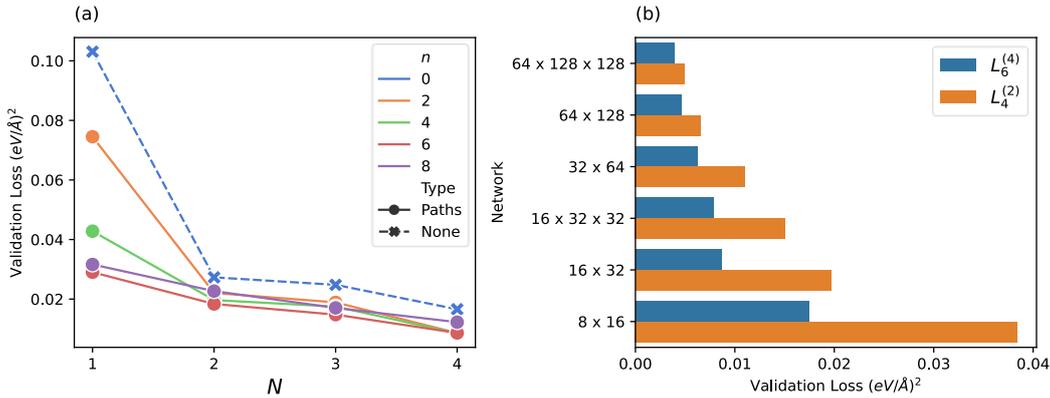

*Figure 1* – (a) The validation loss of the forces prediction for MoO$_3$ plotted as a function of the order of neighbors $N$ for different $n$ values in $L_n^{(N)}$, using the 16×32 BPNNs after 100,000 epochs. (b) The validation loss of the $L_6^{(4)}$- and $L_4^{(2)}$-based models in different network architectures. All networks are dense fully connected with the tanh activation function.

After undergoing an initial training of 100,000 epochs, the $L_6^{(4)}$ and $L_4^{(2)}$-based networks were further trained for an additional 900,000 epochs (Figure 2). This resulted in a final root MSE (RMSE) of 0.056 eV/Å and 0.095 eV/Å (mean absolute error, MAE, of 0.039 eV/Å and 0.069 eV/Å) for the test set, respectively.

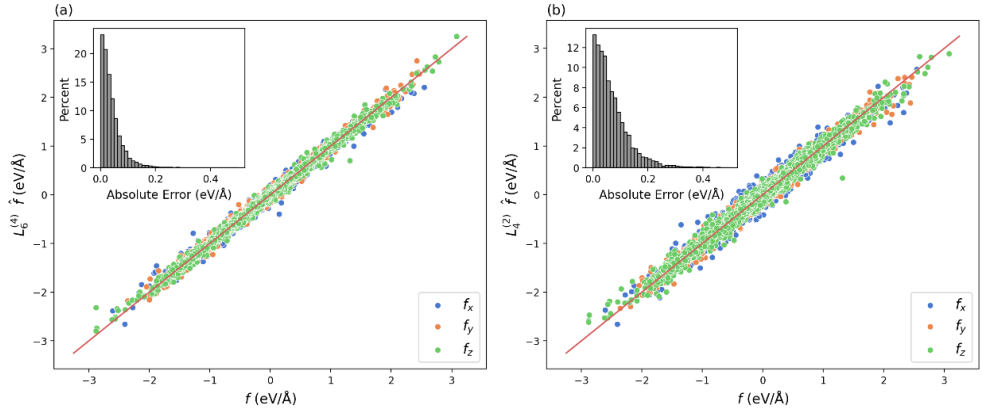

*Figure 2* – Comparison of the calculated and predicted atomic forces of MoO$_3$ for (a) $L_6^{(4)}$ and (b) $L_4^{(2)}$ models, accompanied by insets depicting the absolute errors in force prediction.

These performance metrics are regarded as good for deep potentials application [14] and are comparable to those achieved in previous deep potential work (Figure S5), with the MAE of the $L_6^{(4)}$ and $L_4^{(2)}$ models accounting for less than 10% and 15% of the average atomic forces, respectively.

Next, we apply the same methodology for the BaTiO$_3$ dataset. As mentioned above, the $L_n^{(N)}$ vector for the BaTiO$_3$ system consists solely of simple cycles due to its higher connectivity. Additionally, the higher complexity of the BaTiO$_3$ system requires longer training compared to that of MoO$_3$ (1,000,000 epochs). Thus, we train simple cycles-based $L_n^{(N)}$ models with different $N$ and $n$, using the 16×32 BPNN architecture. The performance benchmark (Figure S6) of the models with different $N$ and $n$ shows a similar trend to that of MoO$_3$, where the addition of both $N$ and $n$ decreases the loss, with $N$ having a greater impact. We choose to continue to train the $L_6^{(4)}$ and $L_4^{(2)}$ models for 2,000,000 additional epochs. The final results for the BaTiO$_3$ system are similar in quality to those of MoO$_3$ and can be used for application in atomistic potentials (Figure S7). The RMSE for the $L_6^{(4)}$- and $L_4^{(2)}$-based models are 0.052 eV/Å and 0.082 eV/Å (MAE of 0.035 eV/Å and 0.062 eV/Å), respectively. The performance of the models demonstrates that even the use of only simple cycles allows reaching a similar performance to that of the full possible paths description. The use of simple cycles is necessary due to the high connectivity of the structures that gives rise to many unique paths with repetitions.

While forces prediction is common, fast and accurate electronic structure prediction using deep potentials has not been widely reported. Therefore, we train BPNNs for the prediction of LDOS using our features. Features performance comparison as a function of $n$ and $N$ shows that for LDOS prediction, the main increase in accuracy is achieved by the incorporation of longer paths (i.e., increasing $n$), while adding higher-order neighbors' coordinates (i.e., increasing N) has a weak effect, as can be seen in Figure 3 (a). While using $N = 0$ is feasible, models with $n = 0$ show insufficient performance.

The difference between the results for the forces where both $n$ and $N$ must be increased to achieve higher accuracy and the results for LDOS where only an increase in $n$ is necessary for higher accuracy can be understood based on the moments theorem. The moments of LDOS are fully specified by the hopping matrix elements along the self-returning paths that describe multicenter chemical bonding, so that these features are dominant for predicting LDOS. By contrast, forces on atoms are due to the combined effect of both the bonding interactions captured by the hopping matrix elements along self-returning paths and the two-body short-range repulsion interactions that are specified by atomic coordinates, making both cycles and coordinates necessary for force prediction.

For LDOS prediction, all of the models were trained for 30,000 epochs. For MoO$_3$, in contrast to forces prediction, $L_6^{(4)}$ shows a much smaller improvement in prediction accuracy compared to $L_4^{(2)}$ (Figure 3 (a)). For BaTiO$_3$, a steeper improvement with increasing $n$ is exhibited, which could be related to the use of cycles. We also find a decrease in the performance in the case of BaTiO$_3$ with $N = 4$, which could be explained by the longer training which may be required for this high-dimensional representation. Thus, for both MoO$_3$ and BaTiO$_3$ we use the $L_4^{(2)}$ model for further LDOS results analysis (Figure 3 (b)).

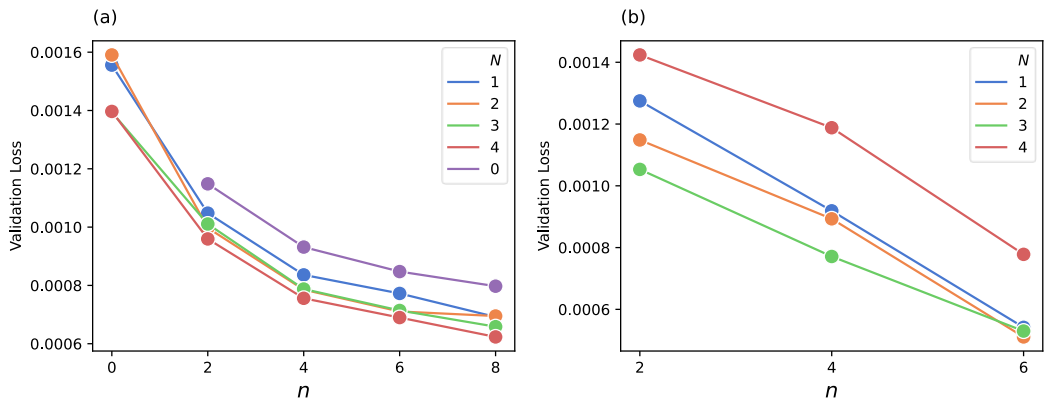

*Figure 3 – (a)* The validation loss in the prediction of LDOS plotted as a function of different $n$ hops for different order of neighbors $N$ in $L_n^{(N)}$, using 16×32 BPNNs after 30,000 epochs for (a) MoO3 and (b) BaTiO$_3$.

Figure *4* shows the real and predicted LDOS for MoO$_3$, for energies close to the Fermi energy, showing good agreement between the model-predicted and calculated LDOS. While it is difficult to compare between different works due to the different nature of the LDOS and variations in the energy ranges which do not contain states, the test error metrics found for both compounds are smaller than or similar to those in previous works [38–40]. However, here LDOS prediction is achieved with a much smaller model.

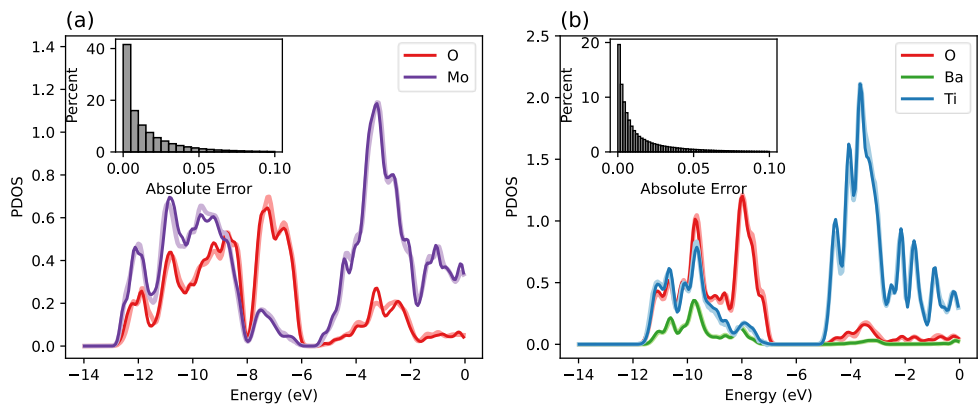

Figure 4 - Real (brighter tone colors) and predicted (darker tone colors) LDOS of a representative (exhibiting the error closest to the MAE) atom in each species in (a) MoO$_3$ and (b) BaTiO3, using $L_4^{(2)}$ and a 16×32 BPNN. Inset shows absolute error distribution of the selected model.

With regard to computational speed, Figure 5 shows a comparison of the time per step of TinyNet to the TIP3P, Bond Valence (BV) [3], and the DPMD models [28]. The combination of the cycles features with small network allows low computational cost for a time step, making the calculation 20 times faster than the DPMD model and only 35% slower than the empirical BV potential. We note that the timings reported in this work include the embedding and forces prediction time for $L_4^{(2)}$ which are implemented in Python. Therefore, it is expected that they will be faster upon implementation in a compiled programming language.

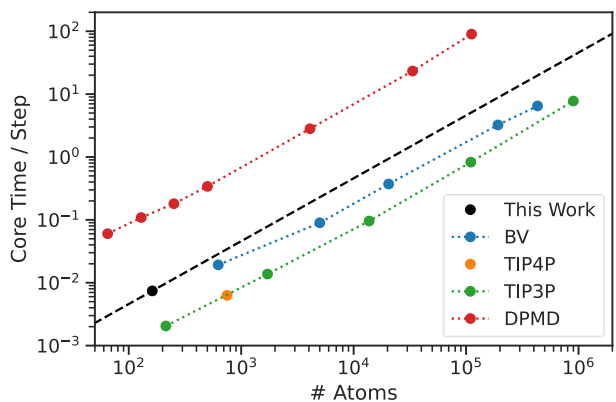

*Figure 5* – Core time per step comparison for different models and the TinyNet $L_4^{(2)}$ model for BaTiO$_3$ suggested in this work, applied for the prediction of forces applied on a 160 atoms supercell. The comparison is based on the timings between those of TIP3P published by Zhang et al [4] and TIP4P in our cluster.

In conclusion, we have developed a chemical-bonding-aware embedding based on the moments theorem for application in atomistic neural network models for MD simulations We demonstrate the use of the embedding for accurate prediction of forces and LDOS for MoO$_3$ and BaTiO$_3$ in a ultrasmall 16x32 BPNN architecture, enabling low computational cost per time step. The utilization of a small network is enabled by the highly descriptive and comprehensive bonding representation provided by the embedding that is directly relevant to the prediction targets. This feature eliminates the need for the network to redundantly learn information that is already well-known to physicists and chemists, thereby facilitating a smoother learning trajectory and easing the overall learning process. The low propagation time step time, coupled with the simplicity of the embedding, allows for short step times and therefore high practicality for applications in MD simulations.

**References**

[1]   M. S. Daw and M. I. Baskes, *Embedded-Atom Method: Derivation and Application to Impurities, Surfaces, and Other Defects in Metals*, Phys Rev B **29**, 6443 (1984).

[2]   M. I. Baskes, *Modified Embedded-Atom Potentials for Cubic Materials and Impurities*, Phys Rev B **46**, 2727 (1992).

[3]   S. Liu, I. Grinberg, H. Takenaka, and A. M. Rappe, *Reinterpretation of the Bond-Valence Model with Bond-Order Formalism: An Improved Bond-Valence-Based Interatomic Potential for PbTiO3*, Phys Rev B Condens Matter Mater Phys **88**, 104102 (2013).

[4]   L. Zhang, J. Han, H. Wang, and R. Car, *Deep Potential Molecular Dynamics: A Scalable Model with the Accuracy of Quantum Mechanics*, Phys Rev Lett **120**, 143001 (2018).

[5]   J. Behler and M. Parrinello, *Generalized Neural-Network Representation of High-Dimensional Potential-Energy Surfaces*, (2007).

[6]   W. Jia, H. Wang, M. Chen, D. Lu, L. Lin, R. Car, and L. Zhang, *Pushing the Limit of Molecular Dynamics with Ab Initio Accuracy to 100 Million Atoms with Machine Learning*, (n.d.).

[7]   R. He, H. Wu, L. Zhang, X. Wang, F. Fu, S. Liu, and Z. Zhong, *Structural Phase Transitions in ${\rm{SrTi}}{{\rm{O}}_3}$ from Deep Potential Molecular Dynamics*, Phys Rev B **105**, 64104 (2022).

[8]   J. Wu, Y. Zhang, L. Zhang, and S. Liu, *Deep Learning of Accurate Force Field of Ferroelectric HfO$_2$*, Phys Rev B **103**, 24108 (2021).

[9]   L. Zhang, D. Y. Lin, H. Wang, R. Car, and E. Weinan, *Active Learning of Uniformly Accurate Interatomic Potentials for Materials Simulation*, Phys Rev Mater **3**, 023804 (2019).

[10]  J. Huang, L. Zhang, H. Wang, J. Zhao, J. Cheng, and E. Weinan, *Deep Potential Generation Scheme and Simulation Protocol for the Li10GeP2S12-Type Superionic Conductors*, Journal of Chemical Physics **154**, 94703 (2021).

[11]  J. Wu, L. Bai, J. Huang, L. Ma, J. Liu, and S. Liu, *Accurate Force Field of Two-Dimensional Ferroelectrics from Deep Learning*, Phys Rev B **104**, 174107 (2021).

[12]  I. A. Balyakin and S. I. Sadovnikov, *Deep Learning Potential for Superionic Phase of Ag 2 S*, (2021).

[13]  D. Guo, C. Li, K. Li, B. Shao, D. Chen, Y. Ma, J. Sun, X. Cao, W. Zeng, and X. Chang, *The Thermoelectric Performance of New Structure SnSe Studied by Quotient Graph and Deep Learning Potential*, Mater Today Energy **20**, 100665 (2021).

[14]  O. Xiu-Juan Li, Z.-B. Zhang, C.-T. Zhang, T. Wen, L. Zhang, H. Wang, and D. J. Srolovitz, *Topical Review Deep Potentials for Materials Science*, (2022).

[15]  D. Dragoni, T. D. Daff, G. Csányi, and N. Marzari, *Achieving DFT Accuracy with a Machine-Learning Interatomic Potential: Thermomechanics and Defects in Bcc Ferromagnetic Iron*, Phys Rev Mater **2**, 013808 (2018).

[16]  J. Behler, *Four Generations of High-Dimensional Neural Network Potentials*, (2021).

[17]  K. T. Schütt, H. E. Sauceda, P. J. Kindermans, A. Tkatchenko, and K. R. Müller, *SchNet - A Deep Learning Architecture for Molecules and Materials*, Journal of Chemical Physics **148**, 241722 (2018).

[18]  K. T. Schütt, P. Kessel, M. Gastegger, K. A. Nicoli, A. Tkatchenko, and K. R. Müller, *SchNetPack: A Deep Learning Toolbox for Atomistic Systems*, J Chem Theory Comput **15**, 448 (2019).

[19]  X. G. Li, C. Hu, C. Chen, Z. Deng, J. Luo, and S. P. Ong, *Quantum-Accurate Spectral Neighbor Analysis Potential Models for Ni-Mo Binary Alloys and Fcc Metals*, Phys Rev B **98**, 094104 (2018).

[20]  C. Chen, Z. Deng, R. Tran, H. Tang, I. H. Chu, and S. P. Ong, *Accurate Force Field for Molybdenum by Machine Learning Large Materials Data*, Phys Rev Mater **1**, 043603 (2017).

[21]  E. V. Podryabinkin and A. V. Shapeev, *Active Learning of Linearly Parametrized Interatomic Potentials*, Comput Mater Sci **140**, 171 (2017).

[22]  V. L. Deringer, A. P. Bartók, N. Bernstein, D. M. Wilkins, M. Ceriotti, and G. Csányi, *Gaussian Process Regression for Materials and Molecules*, Chem Rev **121**, 10073 (2021).

[23]   A. P. Bartók, J. Kermode, N. Bernstein, and G. Csányi, *Machine Learning a General-Purpose Interatomic Potential for Silicon*, Phys Rev X **8**, 041048 (2018).

[24]   A. P. Bartók, M. C. Payne, R. Kondor, and G. Csányi, *Gaussian Approximation Potentials: The Accuracy of Quantum Mechanics, without the Electrons*, Phys Rev Lett **104**, 136403 (2010).

[25]   A. V. Shapeev, *Moment Tensor Potentials: A Class of Systematically Improvable Interatomic Potentials*, Https://Doi.Org/10.1137/15M1054183 **14**, 1153 (2016).

[26]   S. Chmiela, H. E. Sauceda, I. Poltavsky, K. R. Müller, and A. Tkatchenko, *SGDML: Constructing Accurate and Data Efficient Molecular Force Fields Using Machine Learning*, Comput Phys Commun **240**, 38 (2019).

[27]   Y. Zhang, H. Wang, W. Chen, J. Zeng, L. Zhang, H. Wang, and W. E, *DP-GEN: A Concurrent Learning Platform for the Generation of Reliable Deep Learning Based Potential Energy Models*, Comput Phys Commun **253**, 107206 (2020).

[28]   H. Wang, L. Zhang, J. Han, and W. E, *DeePMD-Kit: A Deep Learning Package for Many-Body Potential Energy Representation and Molecular Dynamics*, Comput Phys Commun **228**, 178 (2018).

[29]   J. Behler, *Perspective: Machine Learning Potentials for Atomistic Simulations*, Journal of Chemical Physics **145**, 170901 (2016).

[30]   A. P. Bartók, R. Kondor, and G. Csányi, *On Representing Chemical Environments*, Phys Rev B **87**, 184115 (2013).

[31]   L. Farokhah, T. Herman, A. Jupri -, E. D. Wahyuni, E. Susanti, N. Sari, H. Huo, and M. Rupp, *Unified Representation of Molecules and Crystals for Machine Learning*, Mach Learn Sci Technol **3**, 045017 (2022).

[32]   G. E. Karniadakis, I. G. Kevrekidis, L. Lu, P. Perdikaris, S. Wang, and L. Yang, *Physics-Informed Machine Learning*, Nature Reviews Physics 2021 3:6 **3**, 422 (2021).

[33]   F. Cyrot-Lackmann, *On the Electronic Structure of Liquid Transitional Metals*, Adv Phys **16**, 393 (1967).

[34]   F. Ducastelle and F. Cyrot-Lackmann, *Moments Developments and Their Application to the Electronic Charge Distribution of d Bands*, Journal of Physics and Chemistry of Solids **31**, 1295 (1970).

[35]   J. M. Soler, E. Artacho, J. D. Gale, A. García, J. Junquera, P. Ordejón, and D. Sánchez-Portal, *The SIESTA Method for Ab Initio Order-N Simulation*, Journal of Physics: Condensed Matter **14**, 2745 (2002).

[36]   M. J. van Setten, M. Giantomassi, E. Bousquet, M. J. Verstraete, D. R. Hamann, X. Gonze, and G. M. Rignanese, *The PseudoDojo: Training and Grading a 85 Element Optimized Norm-Conserving Pseudopotential Table*, Comput Phys Commun **226**, 39 (2018).

[37]   M. Abadi et al., *TensorFlow: Large-Scale Machine Learning on Heterogeneous Systems*.

[38]   V. Fung, P. Ganesh, and B. G. Sumpter, *Physically Informed Machine Learning Prediction of Electronic Density of States*, Chem. Mater **2022**, 51 (2022).

[39] P. R. Kaundinya, K. Choudhary, S. R. Kalidindi, and -G 4 W Woodruff, *Prediction of the Electron Density of States for Crystalline Compounds with Atomistic Line Graph Neural Networks (ALIGNN)*, JOM **74**, (n.d.).

[40] N. Lee, H. Noh, S. Kim, D. Hyun, G. S. Na, and C. Park, *PREDICTING DENSITY OF STATES VIA MULTI-MODAL TRANSFORMER*, (n.d.).